# Correlation of the fragility of metallic liquids with the high temperature structure, volume, and cohesive energy

**A. K. Gangopadhyay*, C. E. Pueblo, R. Dai, M. L. Johnson, R. Ashcraft, D. Van Hoesen, M. Sellers, and K. F. Kelton**

Department of Physics and Institute of Materials Science & Engineering, Washington University in St. Louis, St. Louis, MO 63130, USA

*Corresponding author. E-mail: anup@wuphys.wustl.edu

## Abstract

The thermal expansion coefficients, structure factors, and viscosities of *twenty-five* equilibrium and supercooled metallic liquids have been measured using an electrostatic levitation (ESL) facility. The structure factor was measured at the Advanced Photon Source, Argonne, using the ESL. A clear connection between liquid fragility and structural and volumetric changes at high temperatures is established; the observed changes are larger for the more fragile liquids. It is also demonstrated that the fragility of metallic liquids is determined to a large extent by the cohesive energy and is, therefore, predictable. These results are expected to provide useful guidance in the future design of metallic glasses.

## I. Introduction

Amorphous solids (or glasses) are devoid of the long-range order of crystals, but possess significant short- and medium-range order that develops in the supercooled liquids (i.e. below the liquidus temperature, $T_l$) from which they are formed. A remarkable property of liquids is the rapid increase in the magnitude of their dynamical properties (viscosity, diffusion coefficient) on approaching a characteristic temperature, called the glass-transition temperature, $T_g$, where the time-scale for relaxation exceeds the experimental time-scale (typically, 100 seconds). The liquids are classified [1] as "fragile" or "strong" depending on how fast their dynamical properties change near $T_g$. The fragility parameter, defined as $m = (\frac{dlog\eta}{dT})_{T_g}$ (where η is the viscosity), is larger for more fragile liquids. Since the dynamics of a liquid are often considered to be connected to the thermodynamic properties by the Adam-Gibbs relation [2], the fragility parameter can also be determined from thermodynamic quantities, such as specific heat [3] and entropy [4]. The bulk of the existing literature on glasses focus on the dynamical and thermodynamic properties near $T_g$; rarely are high temperature (above and below the liquidus, $T_l$) properties, such as the volume and static structure factor of liquids, used to define fragility. It is demonstrated here that the high temperature properties of equilibrium and supercooled liquids are equally important indicators of liquid fragilities. It is also shown that for the metallic liquids investigated, the cohesive energy plays a significant role in determining liquid fragility. This is a significant new result, demonstrating that the cohesive energy is intimately connected with structural evolution and the dynamical properties of metallic liquids.

It is somewhat surprising that volume and structural properties are not widely used as measures of the liquid fragility, although their role in glass formation is well recognized [5-9].

For example, the dense random packing of hard spheres was one of the earliest models used to describe the structure of metallic glasses [5]. Recently, it has also become clear that the efficient packing of atoms [6] and atomic clusters [7,8] can successfully explain the glass forming abilities of metallic glasses. Such packings produce denser liquids and glasses, and indeed a correlation between density and glass forming ability has been reported in $Cu_{1-x}Zr_x$ [9]. Another study reported a connection between the volume expansion coefficient and glass formability in $Cu_{1-x}Zr_x$ alloys [10]. In both studies, the critical casting thickness of a glass was used as a measure of glass formability. Instead of one single alloy system, we report here a comprehensive study of thermal expansion coefficients for *twenty-five* different metallic liquids made of diverse chemical constituents. The important result is that the volume expansion coefficient correlates well with the liquid fragility, which differs from the earlier reports in two important ways. First, the volume expansion coefficient is larger for the more fragile liquids, which is opposite to the earlier interpretation on a single alloy system [10]. Second, the correlation reported here is not with glass formability, but with liquid fragility. This study is also important for another reason. Although a link between volume and dynamic properties is widely recognized in the context of the free-volume models [11,12], experimental demonstration for such a link between *volume and fragility* in metallic glasses is not well established.

There have been very few studies of structural evolution in high temperature metallic liquids and of its potential connection with fragility and glass formability. Only recently our group has started such investigations, focusing on bulk metallic glasses (BMGs, defined by those that can be cast in more than 1mm thick glasses). We have found that as the height of the first of the static structure factor, $S(q)$, of the liquid is linearly extrapolated to $T_g$, it does not match with that measured in the glass [13-15]. The amount of the mismatch is connected with the liquid fragility. However, a more comprehensive study of a large number of metallic liquids, both marginal and good glass-formers, is still lacking. Focusing exclusively on the high temperature equilibrium and supercooled liquids (above and below $T_l$) we demonstrate here a clear connection between fragility and the rate of change of the liquid structure with temperature. The fragility is manifest in the high temperature liquid, both in the liquid structure factor, $S(q)$, and the pair correlation function, $g(r)$. Unlike our previous study [15], then, there is no need to compare the structural properties with those of the amorphous solids. Taken together, results presented here demonstrate that the fragility of the amorphous solid is a clearly observable property in the volume and structure of the high temperature liquids.

## II. Experimental methods

The density/volume of a large number of liquids that form BMGs and marginal metallic glass-forming liquids were studied in the equilibrium and supercooled states by using the electrostatic levitation (ESL) technique [16-18]. For each composition, a master ingot (about 1g) was prepared by arc-melting high purity (greater than 3N) elements in a high purity (5N) argon-atmosphere, which was improved further by melting a $Ti_{50}Zr_{50}$ getter before each melting. Each ingot was flipped and re-melted four times to ensure homogeneity. Smaller samples (2.2-2.5 mm diameter) for the volume and $S(q)$) measurements were taken from the master ingots and were cast in a nearly spherical shape in a Cu-mold in an arc-melting facility. The sample sizes were limited by the ability to reliably levitate and melt them in the ESL. The density/specific volumes of the levitated samples were obtained from the analysis of two-dimensional video images, as described elsewhere [19-20]. The shapes of the levitated liquids deviated slightly from a perfect spherical symmetry, due to gravitational distortions. Since the distortion was along the vertical axis, however, and the samples were symmetric about that axis, it caused very small error in the measured volume. Additional distortions were often present in the glassy and crystalline samples due to phase transformation from the liquids. However, sample rotation due to radiation pressure from the heating laser averaged out some of this asphericity in the time-averaged video images. To further improve the statistics, the data from each cooling cycle for the glasses were binned over small temperature intervals (10 K) and averaged. Data from several cooling cycles were then averaged to obtain the final results. No such averaging was necessary for the liquids. Typically, volume changes for the glasses could be resolved with an accuracy of 1 part in $10^3$; the typically accuracy for the liquids was one order of magnitude higher (1 part in $10^4$). The absolute precision was limited to about 1 %, determined by the precision of the WC calibration spheres (McMaster-Carr Inc., USA). The lowest measurement temperature for the glasses was about 473 K, set by the pyrometer (Metis, Germany); the highest temperatures were determined by the onset of crystallization. For the liquids, measurements were made from 200-300 K above $T_l$ down to the onset of crystallization in the deeply supercooled liquids (typically, 15-20 % below $T_l$). To check for reproducibility, measurements were made on at least two different samples (more in a few cases) of the same composition.

The structural data for the liquids were obtained at the beam-line 6ID-D at the Advanced Photon Source, Argonne National Laboratory on levitated samples using high energy X-rays (129 keV, 0.0958(6) Å) in a transmission geometry. The scattered intensity was recorded by a GE Revolution 41-RT area detector over a range of $0.8 \leq q \leq 14$ Å$^{-1}$. Background contributions from the Be-window and air scattering were subtracted from the measured intensities. An in-house software package [21], designed to correct for absorption, multiple- and incoherent Compton-scattering, was used to obtain $S(q)$ as a function of temperature.

The oscillating drop method [22] was used for the viscosity measurements. A modulating voltage was applied to the vertical electrodes to induce oscillations of the levitated droplet near the resonance frequency (typically, 90-140 Hz). The perturbing voltage was then removed and the decay of the oscillation was recorded by a high-speed camera (1560 frames per second). The viscosity was determined from the decay time-constant, $\tau$, which is related to viscosity by

$$\eta = \frac{\rho R_o}{(l-1)(2l+1)\tau} \tag{1}$$

where $\rho$ is the density, $R_0$ the sample radius, and $l$ is the order of the harmonic, which was 2 for the present measurements.

The glass transition temperature, $T_g$, was measured with a Perkin-Elmer differential scanning calorimeter (model DSC-8500). Although literature data for $T_g$ exist for most of the glasses studied, different heating rates were used for those measurements and the state of the glasses (as-quenched or relaxed) was often not reported. To be consistent, all measurements reported here were made with a 20 K/min. heating rate for fully relaxed glasses. For those marginal glasses for which no calorimetric signal for $T_g$ were detected, the onset of the first crystallization temperature was approximated as $T_g$. The solidus ($T_s$) and liquidus temperatures ($T_l$) were measured with a differential thermal analyzer (Setaram Labsys).

## III. Results and Discussion

### III.A Fragility and thermal expansion

Figure 1 shows typical data obtained for the specific volume of glassy, crystalline, and liquid phases of a representative BMG. The scatter in the data is larger for the glasses and the crystal phases because of small deviations from a spherical shape, as mentioned earlier. Measurements on the glasses were made during cooling, after they had been fully relaxed by heating to 20-50 K above $T_g$. The data show a linear relationship between volume and temperature within the precision of these measurements. The thermal expansion coefficients of the liquids and glasses were obtained from such data, which are provided in Table 1 along with their $T_g$ and $T_l$.

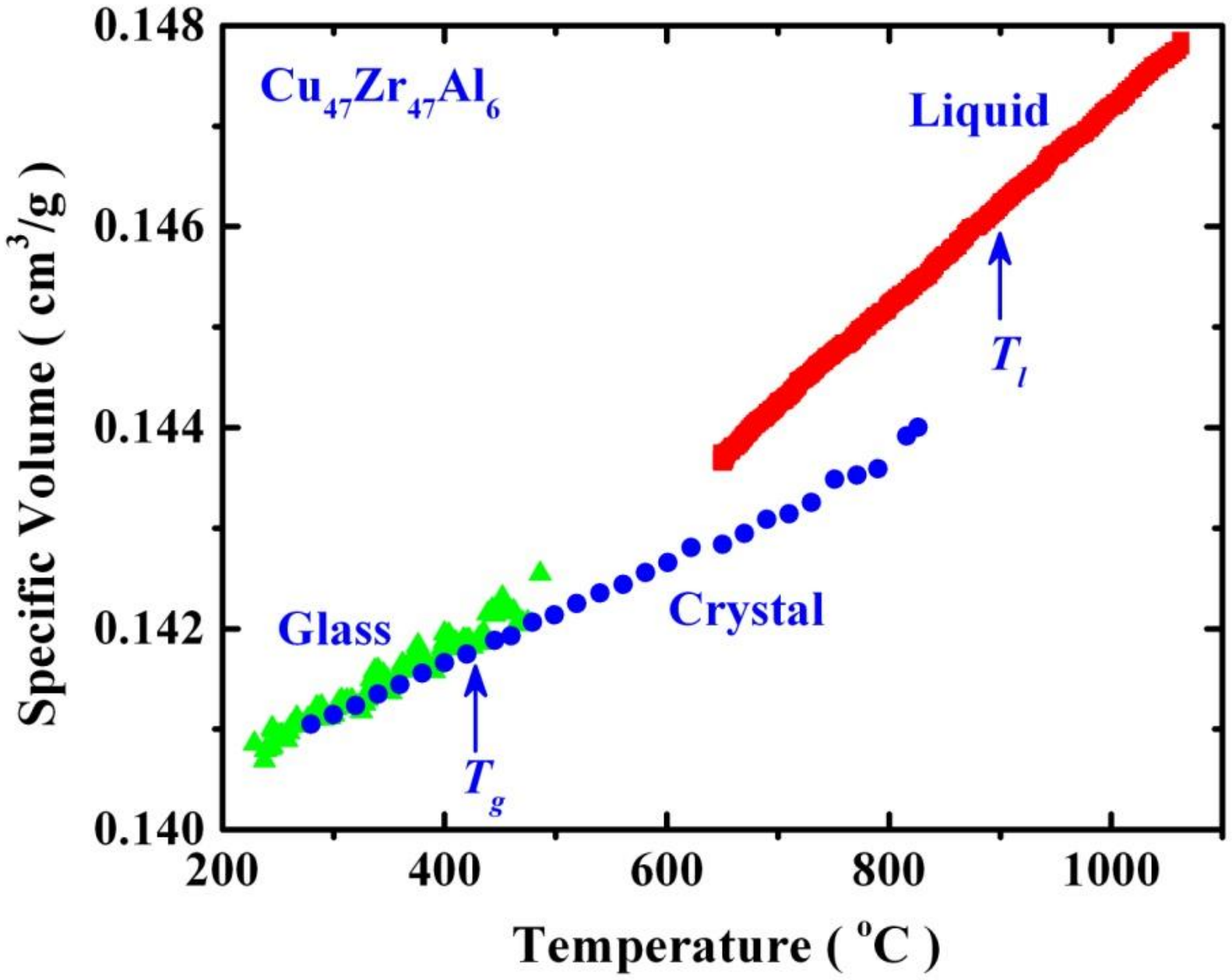

FIG. 1. Specific volumes for the amorphous, crystal, and liquid phases of a $Cu_{47}Zr_{47}Al_6$ alloy (color online).

For crystals and glasses the expansion coefficient is determined primarily by the anharmonicity of the interatomic potential. For liquids additional contributions come from changes in the liquid structure with temperature [23]. To check whether the fragility is linked with the thermal expansion coefficients of the liquids and glasses, or only with the structural contribution to the thermal expansion coefficient, all three parameters were evaluated. To obtain an estimate of the structural contribution from the liquids, the thermal expansion coefficients for the glasses were subtracted from those of the corresponding liquids, since the anharmonic contribution is expected to be similar in the liquids and glasses. This is a general practice followed in the literature for estimates of the configurational entropy of liquids [4], although some dissenting views also exist [24]. To obtain the conventional fragility parameters, viscosity data for the amorphous solids near $T_g$ are required. Since such data for most of the alloys under consideration do not exist and since such measurements cannot be performed using the ESL technique, we introduce a different measure of fragility that can be obtained from the high temperature viscosity data. Moreover, to correlate fragility with the expansion coefficient of the high temperature liquids, it is best to use a measure of fragility determined in that temperature range anyway. The method introduced is based on an earlier suggestion [4].

Figure 2 shows viscosity data for some of the liquids, measured using the ESL technique, as an Angell-plot [1] (log $\eta$ vs. $T_g/T$); similar data for the other alloy liquids under consideration already exist in the literature [15,25]. It is well-known that the viscosity is linear at high temperatures in an Angell plot [1,4] and become non-linear below a crossover temperature [25].

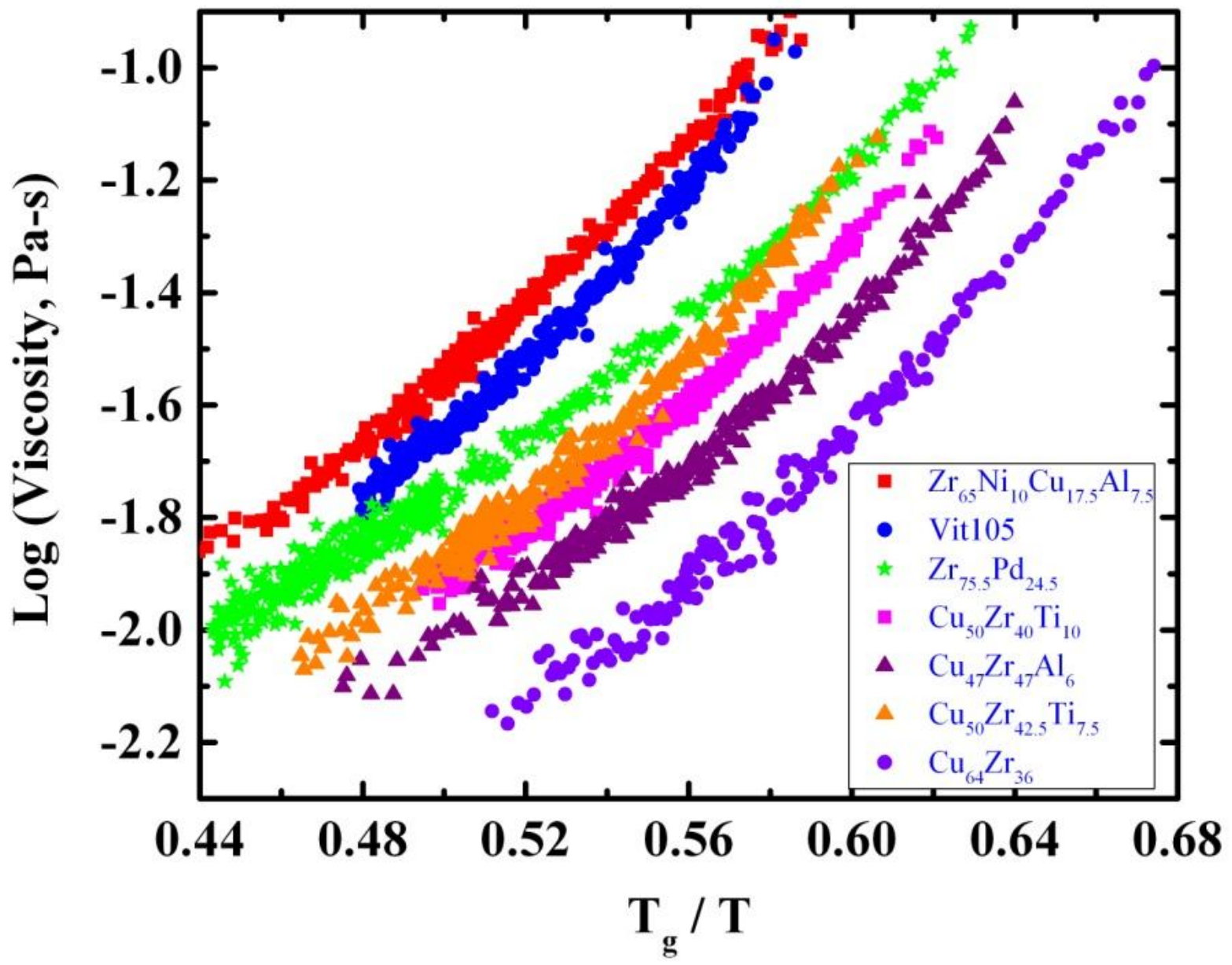

FIG. 2. An Angell-plot of the high-temperature viscosities of some of the liquids under consideration in the present investigation (color online). Data for the other liquids can be found elsewhere [15,25].

Usually, the data for different fragilities do not overlap, as can be seen in Fig. 2. Moreover, the viscosities of stronger liquids are larger than those of the fragile ones at any reduced temperature, $T_g/T$, above $T_g$ [4]. Therefore, for a given value of $\log \eta$ near the liquidus temperature, the corresponding reduced temperature ($T_g/T$) can be used as a measure of fragility. Since data for all liquids under investigation are available at or near a temperature for which $\eta = 0.1$ Pa-s, the inverse of this temperature, $T^*$ (provided in Table 1), scaled to $T_g$ is used as a measure of fragility.

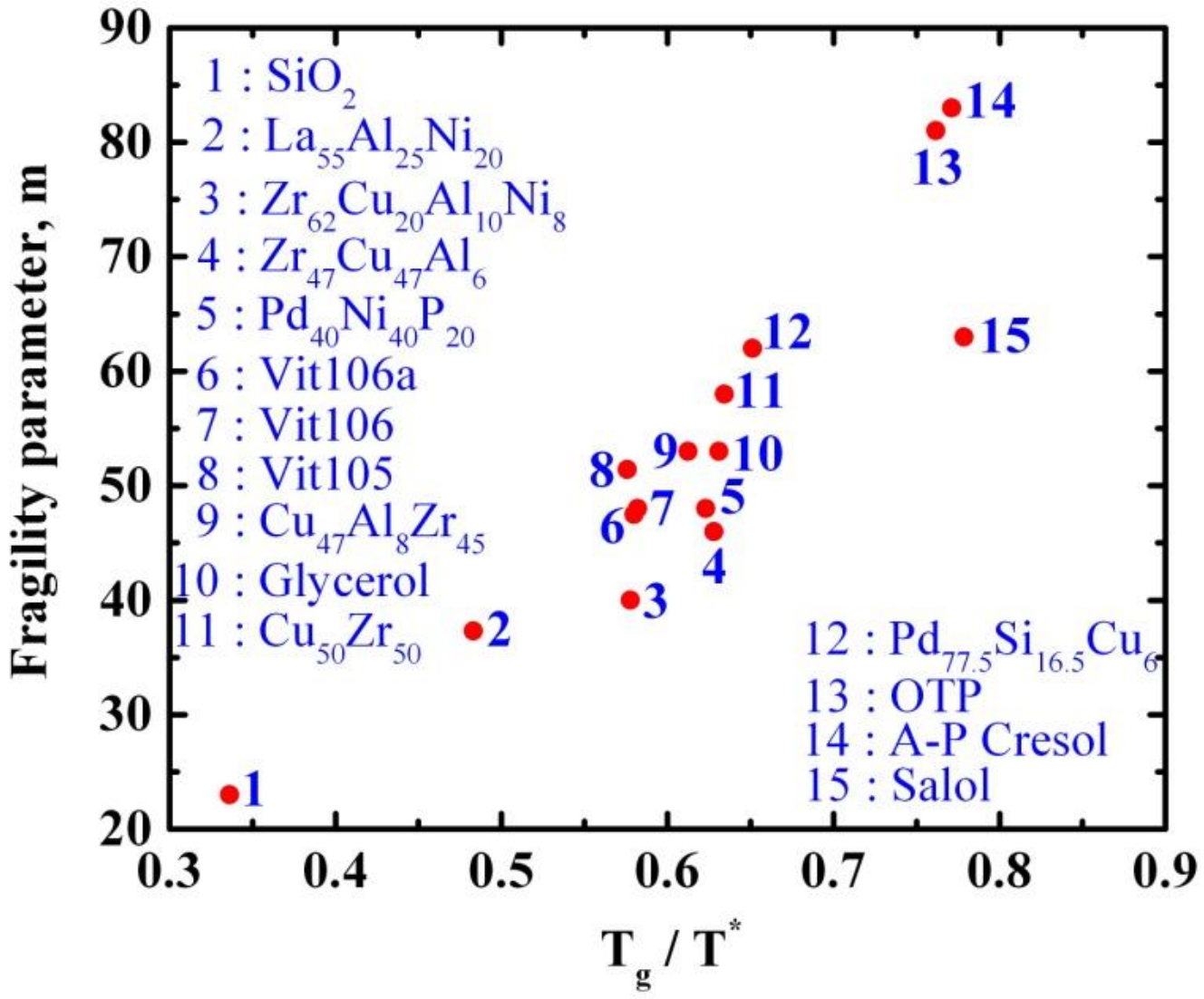

FIG. 3. A comparison of the traditional fragility parameter, $m$, and the high-temperature fragility parameter, $T_g/T^*$, introduced in this investigation (color online). The viscosity is 0.1 Pa-s at $T^*$. The viscosity data used here are taken from refs.: [48] for $SiO_2$, [49] for $La_{55}Al_{25}Ni_{20}$, [50] for $Pd_{40}Ni_{40}P_{20}$, [51] for glycerol, [52] for $Pd_{77.5}Si_{16.5}Cu_6$, and [53] for Ortho-terphenyl (OTP), A-P Cresol, and salol. Data for the other metallic liquids came from this study. The fragility parameter, $m$, for $SiO_2$, glycerol, salol, and OTP are taken from refs. [54], $Cu_{50}Zr_{50}$ from [55], and the metallic glasses from [29].

The choice of $\eta = 0.1$ Pa-s is purely a matter of convenience and no special significance should be attached to this particular value. A different choice for the viscosity would have made no difference for the present discussion, as long as it was kept the same for all liquids. Figure 3 shows the data for $T_g/T^*$ and the traditional fragility parameter, $m$, for some of the metallic liquids for which both types of data are available, including some of the molecular liquids. Clearly, except for Salol, the fragility determined from the high temperature viscosity tracks the conventional fragility parameter, $m$, fairly well. This demonstrates that $T_g/T^*$ is a reliable parameter, and therefore, is used consistently as a measure of fragility throughout this investigation. This result is not surprising, since $T_g/T^*$ parallels the $F_{1/2}$ fragility parameter ($F_{1/2} = 2(T_g/T_{1/2}) - 1$) introduced earlier [4], where $T_{1/2}$ is the temperature at which the logarithm of the viscosity is one half of the difference between the extrapolated infinite temperature viscosity ($\approx 10^{-5}$ Pa-s) and that at $T_g$ ($10^{12}$ Pa-s). The values of $F_{1/2}$ range from about 0.05 for the strongest liquid ($SiO_2$) to 0.8 for one of the most fragile liquids (Ortho-terphenyl) [4]. In comparison, $T_g/T^*$, varies from 0.34 for $SiO_2$ to 0.76 for Ortho-terphenyl.

Figure 4 shows the volume expansion coefficients for metallic liquids as a function of $T_g/T^*$. Note that the data include many of the best glass-formers (*e.g.* Vit 105, Vit106, $Zr_{60}Ni_{25}Al_{15}$) as well as poor glass-forming binary alloy liquids. Except for the two Zr-Pd liquids (nos. 11,12), a clear correlation between the liquid thermal expansion coefficient and fragility is observed. One of the possible reasons for these two outliers may be the uncertainties in determining their $T_g$s. As mentioned earlier, in the absence of a clear calorimetric signal for $T_g$, the onset of the crystallization temperatures were substituted for $T_g$. Otherwise, the data clearly demonstrate that the thermal expansion coefficients are smaller for the stronger liquids (smaller $T_g/T^*$) and larger for the more fragile ones. No such correlation exists, however, between the glass expansion coefficient and fragility. The present results are in contradiction with an earlier study [10] on a single alloy system, where the liquid expansion coefficients were observed to be larger for the better glass formers and stronger liquids [26]. The reason for this discrepancy is not clear at the moment. It follows from the present results that the volume change from the high temperature liquid to the glass (*i.e.* volume expansion coefficient times $T_l - T_g$) is smaller for stronger liquids. Based on the literature data, an earlier study argued for a weak correlation between the critical cooling rate (*i.e.* glass formability) and the volume change in the supercooled liquid between $T_l$ and $T_g$ [27]. The present results show that instead of glass formability, the real correlation is between fragility and the thermal expansion coefficient. As recently demonstrated, fragility is only one parameter that influences glass forming ability [28,29]. The correlation of the expansion coefficient of the liquid and its dynamical fragility is, therefore, a more fundamental connection than one to glass formation alone.

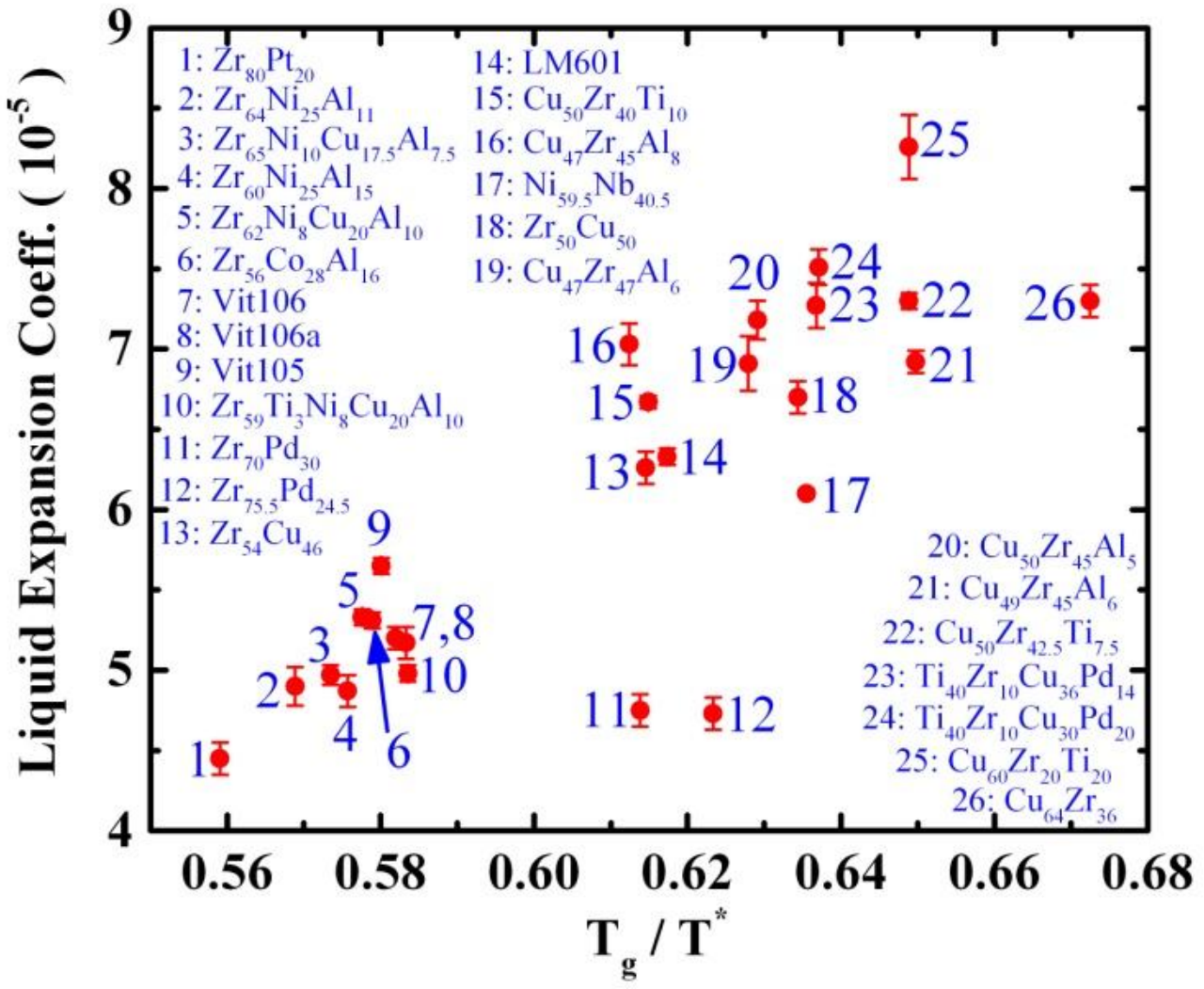

FIG. 4. Correlation between the thermal expansion coefficients for the metallic liquids and the liquid fragility, $T_g/T^*$ (color online). The correlation does not depend on the glass forming ability of the liquids.

It should be pointed out that if the difference between the liquid and glass expansion coefficients is used as a measure of structural contribution to volume expansion, a somewhat similar correlation with the liquid fragility is observed. Such an analysis is not shown here because of greater scatter (larger error in glass expansion coefficients due to sample distortion, see Fig. 1) and limited data points, since only BMGs with larger than 3 mm critical thickness have the appropriate size for the ESL studies. This trend may be verified from the data presented in Table 1. Based on this observation, it is not clear whether the fragility is connected with the total expansion coefficient or only to the structural contribution to the expansion coefficient for the liquids.

**III.B Fragility and liquid structure**

The correlation between volume and fragility suggests that a similar correlation should also hold for the structural changes in supercooled liquids. As mentioned in the introduction, an earlier study [15] focusing on the BMGs demonstrated a correlation between the rapidity of change in the height of the first peak in $S\ q$ near $T_g$ and fragility. It did not, however, consider the rate of structural changes at high temperatures. Therefore, a more detailed study of structural changes in the liquid near the liquidus temperature for a larger number of alloys, including marginal and poor glass-forming liquids, was conducted. The effect of increasing temperature is to increase the disorder of the liquid. In $S(q)$ and $g(r)$ this is manifest by changes in the peak positions, the peak heights, and the peak widths. Therefore, a careful consideration needs to be made to determine which of these changes is most relevant to assess the changes of liquid order. It was shown earlier [30] that the position of the first peak of $g(r)$ with increasing temperature shows an anomalous contraction instead of expansion when the peak is dominated by only one type of partial structure factor. Either contraction or expansion may be observed when several partials contribute significantly to the peak. The positions of higher order peaks always indicate an expansion, but not at rates consistent with the liquid volume expansion coefficients. In comparison, the peak positions for $S(q)$ always indicate an expansion with increasing temperature, as shown earlier [31]. However, the rate of expansion is not what would be expected from the Ehrenfest relation [32]. Similar observations were made in the present investigation for both $g(r)$ and $S(q)$ peaks, which are not shown here for brevity. Due to such complications, the peak positions of $g(r)$ and $S(q)$ are not reliable measures of the changes of order in the liquids.

More reliable metrics for structural changes are the peak heights and peak widths of $g(r)$ and $S(q)$. In figure 5, the heights of the first peak in $g(r)$ for a few representative alloy liquids are shown as a function of temperature. They follow a linear relationship with temperature over the range of measurements, as demonstrated by the straight-line fits. To make the comparison among various alloy compositions meaningful, these rates were normalized by their peak heights at $T_l$. This choice of reference temperature has no fundamental significance for liquid dynamics or ordering, but is a convenient reference since it lies nearly in the middle of the temperature range for which measurements were made. These normalized rates for the first peak heights of $g(r)$ with $T_g/T^*$ are shown in Fig. 6 for *twenty-five* liquids. The rate of change of the peak heights tracks the fragility parameter fairly well; the rates are smaller for the stronger liquids (smaller $T_g/T^*$) and larger for the fragile ones, which are precisely what is observed for the liquid thermal expansion coefficients (Fig. 4). This result does not change if a different reference temperature, other than $T_l$, is chosen. The heights of the first peak in $S(q)$ also show a linear temperature dependence. The normalized rate of change for this peak height is shown in fig. 7 for all liquids. Again a good correlation with the fragility parameter is demonstrated. Instead of the peak heights, if the full-widths at the half-maximum for the first peaks in $g(r)$ and $S(q)$ are considered (not shown here for brevity), their temperature dependences also follow the liquid fragilities. These results clearly demonstrate that the rates of structural changes in the high temperature liquids intimately follow the liquid fragility.

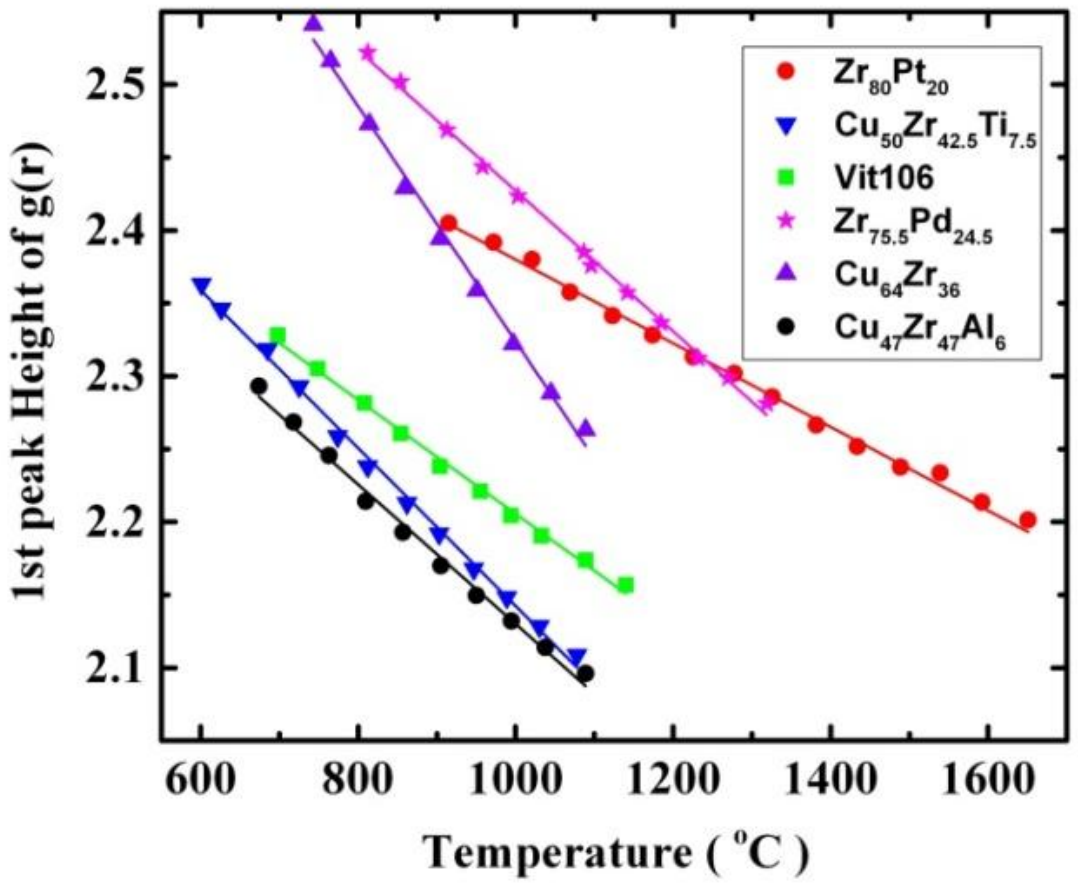

FIG. 5. The temperature dependence of the first peak heights of $g(r)$ for a few representative alloy liquids along with their linear fits (lines) (color online).

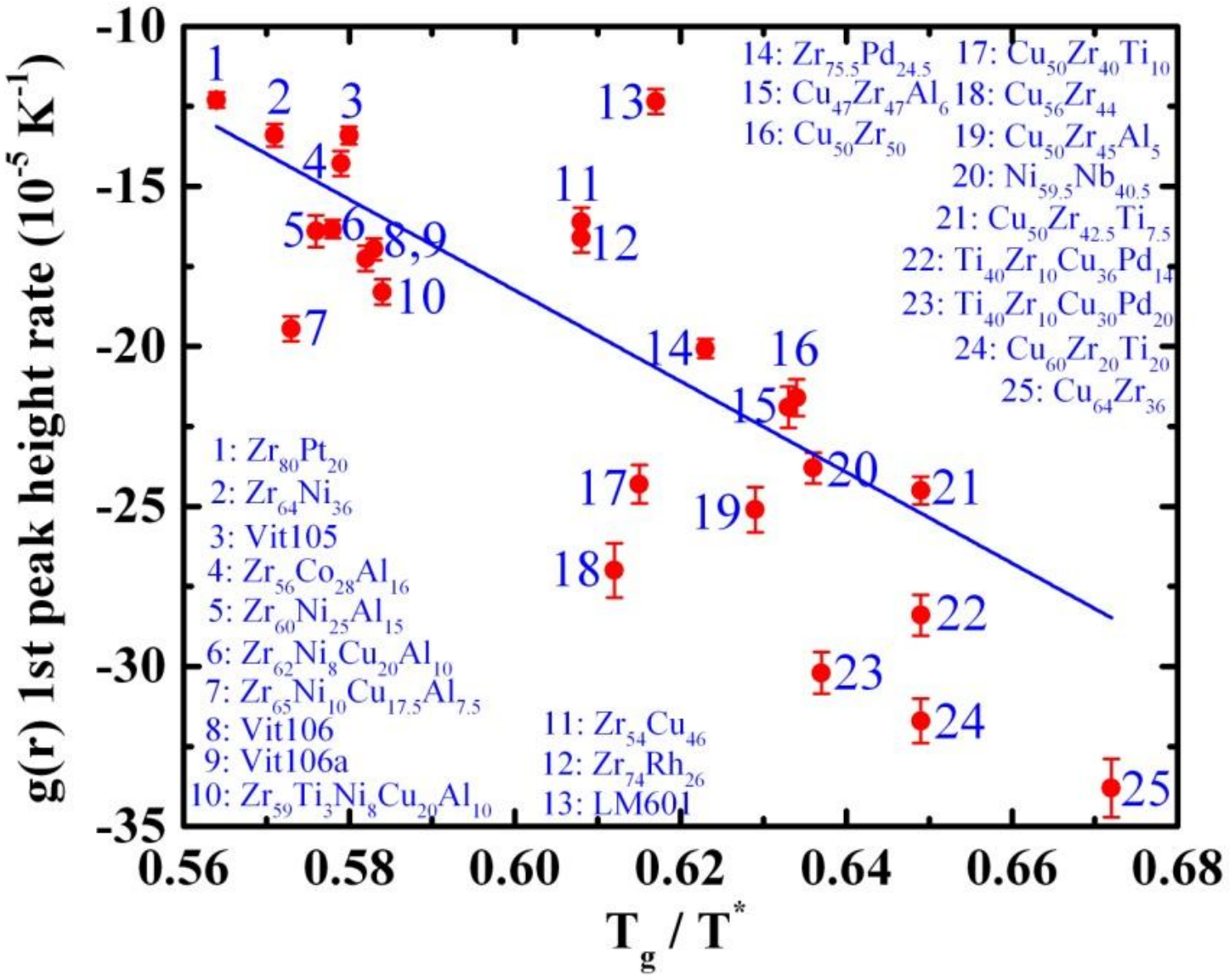

FIG. 6. The normalized rates of change of the first peak height of $g(r)$ with temperature and the liquid fragility as measured by $T_g/T^*$. The straight line is a least-squares fit to the experimental data (color online).

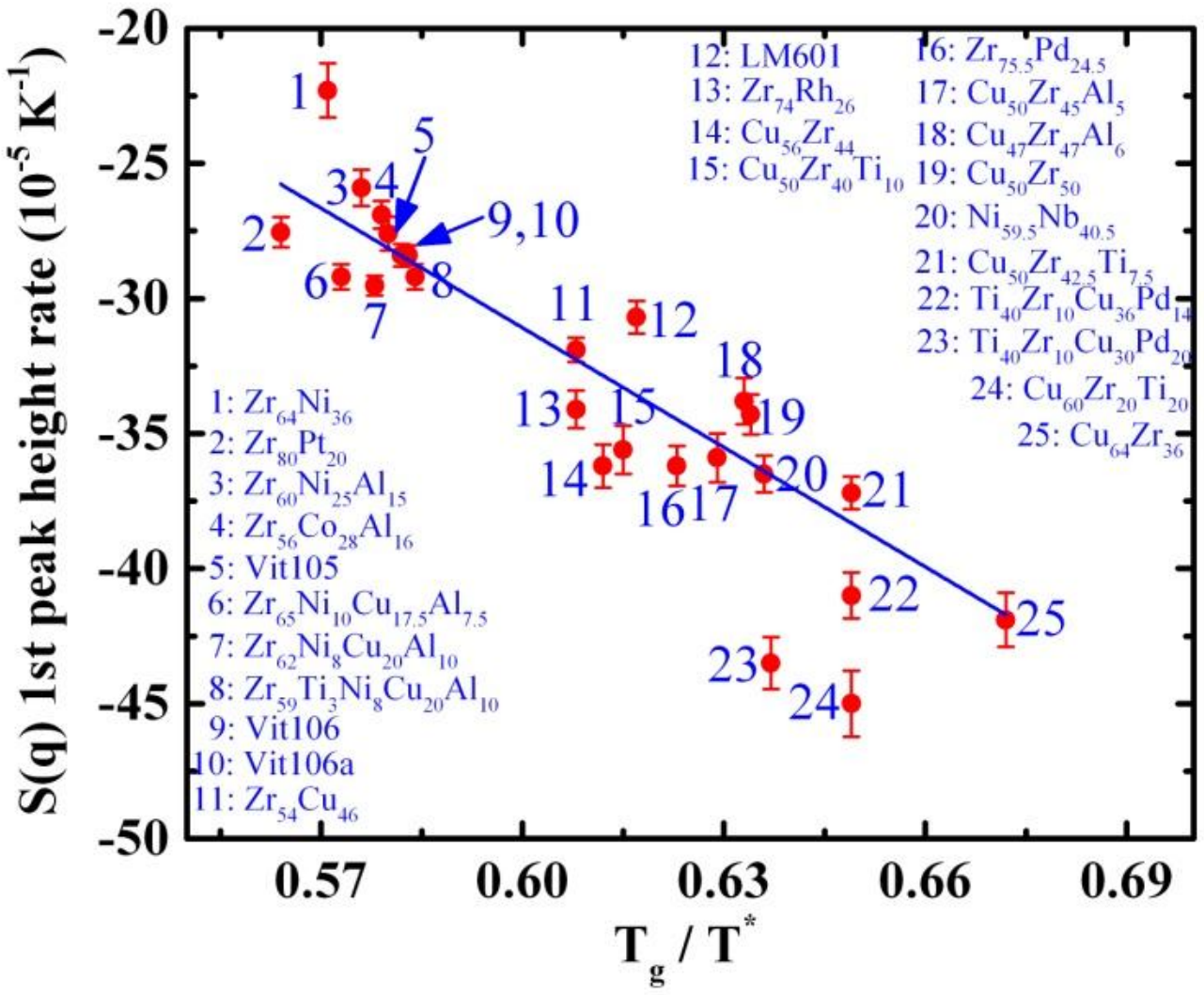

FIG. 7. The normalized rates of change of the first peak height of $S(q)$ with temperature and the liquid fragility as measured by $T_g/T^*$. The straight line is a least-squares fit to the experimental data (color online).

The regression coefficients for the linear fits are, $R^2$=0.67 and 0.76 for the first peaks of $g(r)$ and $S(q)$, respectively. A slightly better fit for the $S(q)$ data may not be that significant considering that the $g(r)$, which comes from the Fourier transformation of the experimental data, $S(q)$, is often plagued by ripples due to truncation error. This may sometime affect the $g(r)$ peak heights. Second-order polynomials, instead of linear fits, also gave similar regression coefficients. Therefore, the functional relationship between fragility and the rates of change of the peak heights is not clear. Whatever the exact functional form may be, the results are clear demonstration that the structure and dynamics are intimately connected. It is important to mention that the heights of the higher order peaks (not peak positions) of $g(r)$ and $S(q)$ also track fragility; but the correlations become progressively weaker (smaller $R^2$) with increasing order of the peaks.

**III.C Fragility and cohesive energy**

Having established a clear connection between structure, volume, and fragility, the microscopic origin of liquid fragility is now considered. An earlier study established a connection between the cohesive energy and thermal expansion coefficients of metallic liquids [33]. In the context of the present results, then, it is natural to ask whether the cohesive energy underlies liquid fragility. Fragility has been linked with the character of the atomic interaction potential [34,35] and the ratio of the infinite frequency bulk modulus and shear modulus of the liquids [36,37]. Since these are all dependent on the interaction potential, a presumed connection between cohesive energy and fragility is reasonable. To check this, the cohesive energies of alloy liquids were estimated from the cohesive energies of the elemental solids [38] and the heats of mixing of the elements [39], following an earlier suggestion [33]. The cohesive energy for a binary alloy, $A_{1-x}B_x$, was estimated using the relationship:

$$E_c = \ 1 - x \ E_A + xE_B + E_{AB} \tag{2}$$

where $E_A$ and $E_B$ are the cohesive energies of the elements A and B and $E_{AB}$ is the heat of mixing for the alloy composition $A_{1-x}B_x$. For a ternary alloy, $A_{1-x-y}B_xC_y$, if the heat of mixing between the solute atoms is small compared with the solvent atoms, then it is effectively a pseudo-binary alloy; the solvent atoms are proportionally distributed with the solute atoms to form $x \ (x+y)$ moles of alloy $A_{1-x-y}B_{x+y}$ and $y \ (x+y)$ moles of alloy $A_{1-x-y}C_{x+y}$. Accordingly, the cohesive energy can be estimated using an equation similar to eqn. 2. This should be a suitable approximation for the binary and pseudo-ternary alloys, but is not expected to be true for multi-component alloys, since the heats of mixing are likely to be determined by more complicated interactions among the various constituents. Due to the absence of a reliable method to estimate the heats of mixing for such alloys, only simple binary and pseudo ternary alloys were considered for this analysis. The estimated cohesive energies are reported in Table 1.

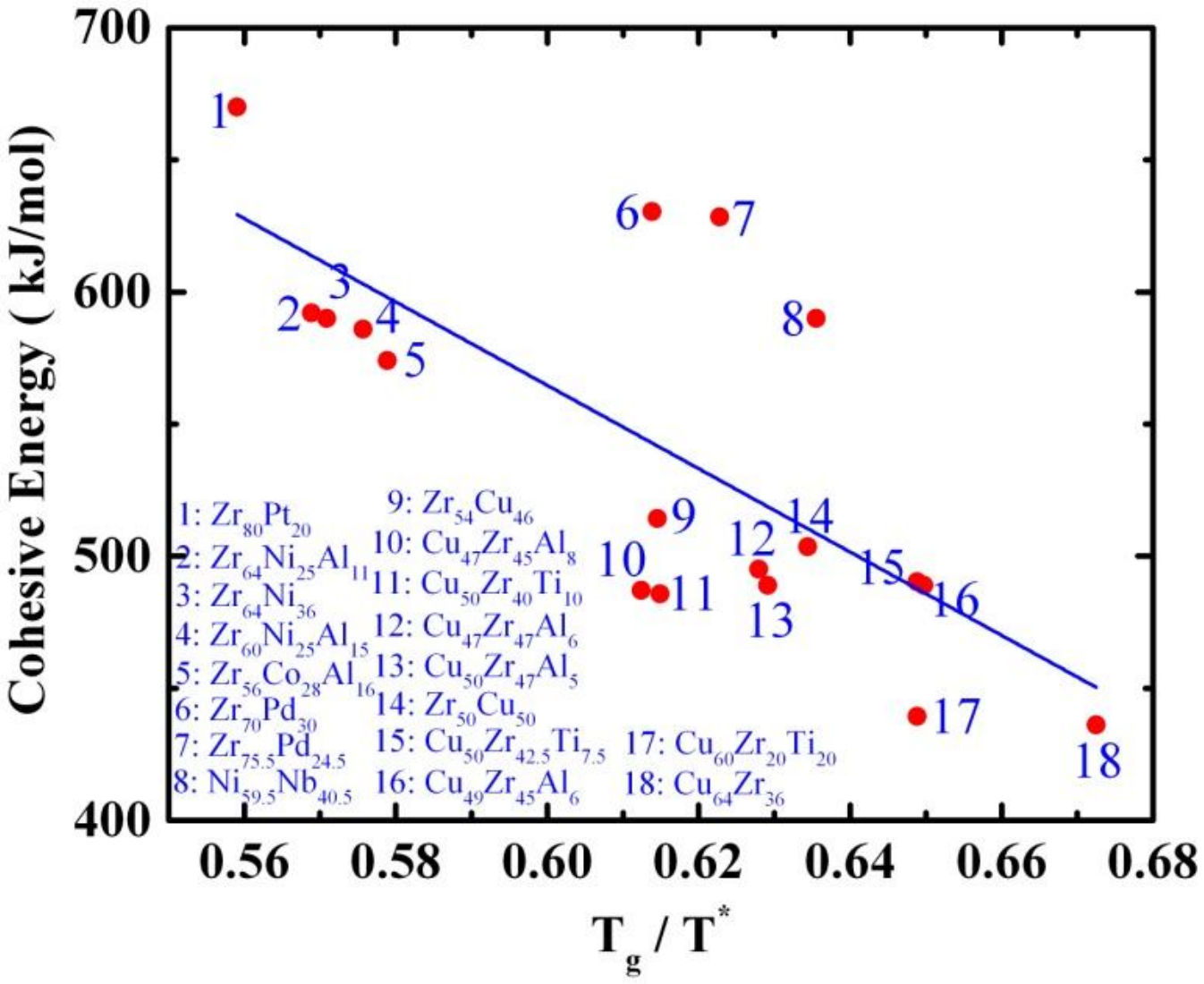

Fig. 8. Correlation between estimated cohesive energy and fragility for binary and ternary liquids (color online).

Figure 8 shows the results from this analysis for *eighteen* binary and ternary liquids based on Zr, Cu, and Ni. Considering the simplicity of the method, a remarkable correlation between fragility and cohesive energy is observed; the larger the cohesive energies, the stronger are the liquids. While the correlation is good, it is not perfect; the regression coefficient ($R^2$) is 0.50, which is not unreasonable, considering the approximations involved. Several sources likely contribute to the observed scatter. The two Zr-Pd (nos. 6 and 7) and the $Ni_{59.5}Nb_{40.5}$ (no. 8) alloys show the maximum deviations from the straight-line fit. If those three data points are removed, the $R^2$ improves significantly to 0.70. Likely, approximating the crystallization onset temperatures as $T_g$ for these two alloys is an overestimation; the actual $T_g$ may be lower than that, which would bring the two data points closer to the straight line fit. The same might be true for the $Ni_{59.5}Nb_{40.5}$ alloy. The other possible reason is that the cohesive energies are estimated at $0^oK$, whereas the experimental data are near $T_l$. Given that the specific heats of liquids are about 25-30 J/mol, the enthalpy correction to the cohesive energy at $T_l$ is likely to be about 25-30 kJ/mol, which is about 5-7% of the estimated cohesive energy at $0^oK$. Interestingly, the typical scatter of the data (except for the alloys 6-8) is also of that order. Leaving aside such details, the data demonstrate that the cohesive energy is a significant contributor to the fragility in metallic liquids.

Perhaps, this relationship is not too surprising, given that the traditional measure of fragility, $m$, is proportional to the ratios of the high frequency (instantaneous) bulk and shear moduli [36]. If the shear moduli for the liquids are not too different, then it is quite natural to expect a direct relationship between fragility and bulk modulus. Although the experimental data for these properties do not exist, the shear moduli for some of the Cu- [40] and Zr-based [41] glasses are quite similar. Since the bulk moduli and cohesive energies are intimately related [38], a connection between fragility and cohesive energy is expected. The important implication of the present result is that, to a first approximation, it is possible to predict the fragility of metallic liquids from the cohesive energy without any experimental parameter.

### III.D Fragility and Glass-formability

Given the significance of fragility as a deciding factor for volume and structural changes in metallic liquids as demonstrated in this work, a few points about fragility and glass formability should be emphasized. While it is widely believed among the metallic glass community that stronger liquids are better glass formers [42-44], the data presented in figures 4, 6, 7, and 8 clearly demonstrate that this is not necessarily the case. The alloys investigated include many binary marginal glass formers, which show comparable fragility to many of the best glass-formers. For example, $Zr_{80}Pt_{20}$ is the strongest liquid among all of the compositions studied in this investigation. In contrast, it is one of the poorest glass formers, producing nano-quaiscrystals even with melt-quenching [45]. Similarly, the binary Zr-Ni alloys are marginal glass formers that are amorphous only as thin-ribbons, produced by conventional melt-quenching technique [46]. Yet, the fragilities of these liquids are comparable to those of $Zr_{60}Ni_{25}Al_{10}$ and $Zr_{64}Ni_{25}Al_{11}$, which are some of the best glass formers with critical thicknesses of 15 and 12 mm respectively [47]. Similarly, the critical casting thickness of $C_{46}Zr_{54}$ is only 0.8 mm [9], but its fragility is comparable to that of LM601, which is a much better glass former. Finally, the critical casting thickness of $Cu_{64}Zr_{36}$ (1 mm) is comparable to $C_{46}Zr_{54}$ (0.8 mm) [9], although the former is much more fragile than the latter. These observations show that taken by itself, fragility is only a weak indicator for good glass formation. Using literature data, Johnson *et al.* [29] recently showed that the combination of the reduced glass temperature, $T_g/T_l$, and the fragility give a good prediction of glass formability. Similar results, which will be presented elsewhere, were obtained in the present investigation for the alloys for which the critical thicknesses are known. The present study, therefore, extends and reinforces those observations from a completely different set of experiments and materials.

## IV. Conclusion

In summary, a clear connection is established between fragility, volume, and structure from, perhaps, the most comprehensive study on metallic liquids. It demonstrates that estimates for the fragility parameter can be made from the volume and structure factor of the high temperature liquids. These results are particularly important for fragile liquids, since the traditional fragility parameters are difficult, or impossible, to measure because of rapid crystallization near $T_g$; properties of equilibrium liquids (above $T_l$) are not restricted by such kinetic constraints. Finally, since the fragility of the metallic liquids appears to be determined predominantly by the cohesive energy of the liquid, estimates for fragility can be made even without experimental measurements. When combined with the reduced glass temperature, which can be estimated from high temperature liquid viscosity data [25], a new approach for the design of metallic glasses with desired properties is achieved.

**Acknowledgments**

This work was partially supported by NASA under grants NNX10AU19G and NNX16AB52G and the NSF under grants DMR 15-06553 and DMR 12-06707. The synchrotron measurements were made on the Sector 6 beamline at the Advanced Photon Source. The technical support provided by Dr. D. S. Robinson and S. Chen during the scattering experiments is gratefully acknowledged. Use of the Advanced Photon Source is supported by the US Department of Energy, Basic Energy Science, Office of Science, under contract no. DE-AC02-06CH11357. Any opinions, findings, and conclusions or recommendations expressed in this material are those of the author(s) and do not necessarily reflect the views of the National Science Foundation or of NASA.

## Table 1

The measured glass transition temperature, liquidus, volume expansion coefficients of the liquids and glasses, and the estimated cohesive energies of the metallic alloys studied here.

| Alloy Composition | Glass Transition Temperature ($T_g$) in K | Liquidus Temperature ($T_l$) in K | Liquid Expansion Coefficient ($10^{-5}$) | Glass Expansion Coefficient ($10^{-5}$) | $T^*$ (K) Temperature for 0.1 Pa-s viscosity | Cohesive Energy (kJ/mol) |
|---|---|---|---|---|---|---|
| $Cu_{64}Zr_{36}$ | 733 | 1200 | 8.43(0.15) | | 1090 | 436 |
| $Cu_{50}Zr_{50}$ | 675 | 1222 | 6.70(0.15) | | 1064 | 503 |
| $Cu_{46}Zr_{54}$ | 633 | 1198 | 6.26(0.20 | | 1030 | 514 |
| $Cu_{50}Al_5Zr_{45}$ | 687 | 1173 | 7.18(0.12) | 4.07(0.23) | 1092 | 489 |
| $Cu_{49}Al_6Zr_{45}$ | 705 | 1177 | 6.92(0.07) | 3.97(0.15) | 1085 | 489 |
| $Cu_{47}Zr_{47}Al_6$ | 697 | 1172 | 6.91(0.17) | 4.05(0.1) | 1110 | 495 |
| $Cu_{47}Al_8Zr_{45}$ | 681 | 1163 | 7.03(0.13) | 3.77(0.17) | 1112 | 487 |
| $Cu_{50}Zr_{40}Ti_{10}$ | 661 | 1168 | 6.67(0.04) | | 1075 | 486 |
| $Cu_{50}Zr_{42.5}Ti_{7.5}$ | 669 | 1152 | 7.3(0.05) | | 1031 | 490 |
| $Cu_{60}Zr_{20}Ti_{20}$ | 693 | 1127 | 8.26(0.2) | | 1068 | 439 |
| $Ti_{40}Zr_{10}Cu_{36}Pd_{14}$ | 675 | 1185 | 7.27(0.14) | 3.80(0.1) | 1060 | |
| $Ti_{40}Zr_{10}Cu_{30}Pd_{20}$ | 690 | 1189 | 7.51(0.11) | 4.20(0.15) | 1083 | |
| $Zr_{64}Ni_{36}$ | 628 | 1283 | 4.89(0.2) | | 1083 | 600 |
| $Zr_{70}Pd_{30}$ | 701 | 1350 | 4.65(0.1) | | 1142 | 630 |
| $Zr_{75.5}Pd_{24.5}$ | 700 | 1303 | 4.73(0.1) | | 1123 | 628 |
| $Zr_{80}Pt_{20}$ | 710 | 1450 | 4.60(0.1) | | 1270 | 670 |
| $Zr_{60}Ni_{25}Al_{15}$ | 692 | 1248 | 4.87(0.1) | 2.94(0.06) | 1202 | 586 |
| $Zr_{64}Ni_{25}Al_{11}$ | 669 | 1212 | 4.90(0.12) | 3.60(0.1) | 1176 | 592 |
| $Zr_{65}Cu_{17.5}Al_{7.5}Ni_{10}$ | 640 | 1170 | 4.97(0.06) | 3.06(0.1) | 1116 | |
| Vit105($Zr_{52.5}Ti_5Cu_{17.9}Ni_{14.6}Al_{10}$) | 670 | 1093 | 5.65(0.05) | 3.25(0.18) | 1155 | |
| Vit106($Zr_{57}Nb_5Cu_{15.4}Ni_{12.6}Al_{10}$) | 678 | 1123 | 5.20(0.07) | 3.12(0.1) | 1165 | |
| Vit106a($Zr_{58.5}Nb_{2.8}Cu_{15.6}Ni_{12.8}Al_{10.3}$) | 672 | 1125 | 5.17(0.1) | 3.06(0.2) | 1152 | |
| LM601($Ti_{51}Cu_{36}Ni_4Al_9$) | 697 | 1157 | 6.33(0.05) | | 1129 | |
| $Zr_{62}Ni_8Cu_{20}Al_{10}$ | 655 | 1152 | 5.33(0.05) | 2.95(0.3) | 1134 | |
| $Zr_{59}Ti_3Ni_8Cu_{20}Al_{10}$ | 653 | 1145 | 4.98(0.1) | 2.99(0.07) | 1119 | |
| $Zr_{56}Co_{28}Al_{16}$ | 730 | 1241 | 5.31(0.05) | | 1210 | 574 |
| $Ni_{59.5}Nb_{40.5}$ | 891 | 1448 | 6.10 | | 1414 | 590 |